\documentclass{article}

\usepackage{arxiv}
\usepackage[utf8]{inputenc} % allow utf-8 input
\usepackage[T1]{fontenc}    % use 8-bit T1 fonts
\usepackage{hyperref}       % hyperlinks
\usepackage{url}            % simple URL typesetting
\usepackage{booktabs}       % professional-quality tables
\usepackage{amsfonts}       % blackboard math symbols
\usepackage{nicefrac}       % compact symbols for 1/2, etc.
\usepackage{microtype}      % microtypography
\usepackage{lipsum}		% Can be removed after putting your text content
\usepackage{graphicx}
\usepackage{natbib}
\usepackage{doi}
\usepackage{multirow}
\usepackage[table,xcdraw]{xcolor}
\usepackage{epstopdf}
\usepackage[english]{babel}
\usepackage{epsfig}
\usepackage{amssymb}
\usepackage{amsmath}

\title{The Use of Alendronate to Enhance Transcranial Transmission of Focused Ultrasound for Successful Ablations in Brain}

\author{ {\hspace{1mm}G. Sakharova} \\
	V.S. Buzaev International Medical Centre\\
	Ufa, Russia \\
	\And
	{\hspace{1mm}A. Krokhmal}\thanks{Corresponding author. Email: krokhmalaa@my.msu.ru}  \\
	Lomonosov Moscow State University\\
	Moscow, Russia\\
    \And
    {\hspace{1mm}R. Galimova} \\
	V.S. Buzaev International Medical Centre\\
	Ufa, Russia \\
    Bashkir State Medical University of the Ministry \\
    of Health of the Russian Federation\\
	Ufa, Russia \\
	\And
    {\hspace{1mm}A. Khatmullina} \\
	V.S. Buzaev International Medical Centre\\
	Ufa, Russia \\
	\And
    {\hspace{1mm}D. Nabiullina} \\
	V.S. Buzaev International Medical Centre\\
	Ufa, Russia \\
	\And
    {\hspace{1mm}I. Buzaev} \\
	V.S. Buzaev International Medical Centre\\
	Ufa, Russia \\
    Bashkir State Medical University of the Ministry \\
    of Health of the Russian Federation\\
	Ufa, Russia \\
	\And
    {\hspace{1mm}D. Avzaletdinova} \\
	Bashkir State Medical University of the Ministry\\
    of Health of the Russian Federation\\
	Ufa, Russia \\
	\And
    \And
	{\hspace{1mm}D. Chupova}\\
	Lomonosov Moscow State University\\
	Moscow, Russia\\
    \And
	{\hspace{1mm}V. Khokhlova} \\
	Lomonosov Moscow State University\\
	Moscow, Russia\\    
	} 

\renewcommand{\shorttitle}{The Use of Alendronate to Enhance tFUS in Brain}

\begin{document}
\maketitle

\begin{abstract}

\textbf{Objective:} The aim of this study was to evaluate the efficacy of alendronate therapy in improving bone density distribution in skull bones and corresponding ultrasound permeability in patients who had previously experienced unsuccessful transcranial MR-guided focused ultrasound (MRgFUS) ablation. The ability of alendronate treatment to modify skull bone characteristics and enhance the success rate of repeat MRgFUS procedures was assessed.

\textbf{Methods:} Five patients with initially unsuccessful MRgFUS ablations underwent a 6–12 month regimen of alendronate to improve bone density. Repeat MRgFUS procedures were performed, and changes in skull density ratio (SDR) and peak focal temperatures were evaluated statistically using CT and MR imaging. Histograms of skull bone density were introduced and analysed as an additional metric.

\textbf{Results:} After therapy, SDR increased in four out of five patients (from 0.378±0.037 to 0.424±0.045, p>0.05). All repeated procedures were successful. The maximum focal temperature, averaged over sonications, increased from 53.6±4.0°C to 55.7±4.1°C (p=0.018), while the maximum temperature per patient rose from 57.0±2.4°C to 60.2±1.8°C (p=0.031). Histograms of CT scans showed a reduction in low-density voxels, indicating trabecular bone densification. 3D CT scan registration revealed local density changes, defect filling, and void reduction.

\textbf{Conclusions:} Alendronate therapy enhanced skull bone density distribution and thus ultrasound permeability, which has facilitated successful repeat MRgFUS. By visually analysing CT changes, healthcare professionals can better inform their decision-making regarding repeat surgeries. This method broadens the pool of patients with low SDR eligible for MRgFUS treatment and underscores the potential benefits of alendronate in improving treatment outcomes.
\end{abstract}

% keywords can be removed
\keywords{MR-guided focused ultrasound \and  FUS \and thermal ablation \and CT images analysis \and Parkinson’s disease \and Essential Tremor \and Dystonia \and trabecular bone \and skull}

\section{Introduction}
\label{sec1}

One of the most promising recent approaches for treating symptoms of movement disorders such as Parkinson’s disease, essential tremor, dystonia, and neurotic pain is thermal ablation using MRI-guided focused ultrasound MRgFUS) \cite{elias2013pilot, lipsman2013mr, martin2009high, galimova2023mri, martinez2018focused}. This method offers several clinical advantages, including non-invasive targeting of deep brain structures, lower postoperative risks compared to deep brain stimulation, reversibility of effects in the early stages of the procedure, and a painless experience without the need for anesthesia during ultrasound exposure (sonication). MRgFUS is often used for patients who do not respond to drug therapy.

A key requirement for selecting patients for MRgFUS treatment is sufficient ultrasound permeability of the skull. The skull acts as a barrier to ultrasound waves by reflecting, scattering, and absorbing its energy, preventing the target tissue from reaching the necessary temperature for effective ablation (for typical MRgFUS sonication times, thermal ablation of brain tissue occurs if the peak temperature reached 55°C \cite{mcdannold2004mri}). Increasing ultrasound power to achieve this target temperature can lead to excessive heating of the skull, posing a serious risk of damage to healthy brain regions and serving as a contraindication for continuing the procedure. 

To address this challenge, the skull density ratio (SDR) was introduced as a key criterion for patient eligibility \cite{gagliardo2020transcranial}. SDR is defined as the ratio of the minimum to maximum skull density along the ultrasound beam paths from the transducer to the focal point, averaged over all beams \cite{chang2016factors}. It is calculated from CT images of the skull, as there is a well-established linear relationship between Hounsfield units (HU) and bone density \cite{schneider1996calibration}. A commonly accepted values for MRgFUS eligibility are SDR > 0.4 \cite{wang2018transcranial},as these values are associated with effective ultrasound penetration and successful heating at the target site  \cite{d2019impact, boutet2019relevance}. Here, the threshold of the SDR value of 0.35 was used for the initial eligibility criterion. 

However, recent studies suggest that SDR is not a universally reliable predictor of treatment success \cite{boutet2019relevance, tsai2021distribution}. There have been cases in which procedures were successful despite SDR values below 0.35, as well as cases where treatment failed despite high SDR values \cite{vetkas2023successful, d2019impact, hino2024effectiveness, ng2024magnetic}. Unsuccessful results are typically due to one of two factors: the target temperature did not reach the required threshold even at maximum ultrasound power, or the patient experienced severe pain during sonication due to excessive skull heating \cite{ng2024magnetic}.  

Both issues may arise from high skull density contrast and structural inhomogeneity \cite{chang2016factors, krokhmal2025comparative}. Variations in geometry and  thickness  of the skull cannot be altered by drug therapy but they can be effectively compensated using aberration correction methods \cite{aubry2003experimental, rosnitskiy2019simulation, hynynen2010mri, chupova2024compensation}. In turn, studies have shown that bisphosphonates can increase bone mass and uniformity of the bone structure by reinforcing the porous regions of the skull \cite{rodan1997bone}. Bisphosphonates are a class of drugs used to treat osteoporosis, reduce the risk of fractures, and manage conditions such as hypercalcaemia and bone tumours (e.g., multiple myeloma and metastatic cancer) \cite{fleisch2002bisphosphonates}. These drugs work by inhibiting osteoclast activity, thereby preserving bone mass.  

Alendronate, a commonly used bisphosphonate, specifically inhibits osteoclastic bone resorption without affecting bone formation \cite{porras1999pharmacokinetics, roschger1997mineralization}. This shifts the balance in favour of bone formation, leading to increased degree and uniformity of mineralization in bone \cite{roschger2001alendronate}. Alendronate treatment has been shown to reduce skull porosity, primarily in the trabecular layer \cite{chavassieux1997histomorphometric}, by enhancing the uniformity of mineralization in the bone matrix, thereby improving structural integrity \cite{roschger2001alendronate}. 

Reduced porosity improves ultrasound conductivity, which often translates to higher SDR values, potentially eliminating one of the key limitations for MR-guided FUS eligibility.  
Several studies have reported increases in SDR in patients who took bisphosphonates for three months, suggesting that this approach could expand access to MRgFUS treatment for patients who would otherwise be ineligible due to poor skull ultrasound permeability \cite{yamamoto2019efficacy}.  In our study, we analysed five clinical cases of patients with SDR ranging from 0.32 to 0.42, in whom focused ultrasound surgery was unsuccessful. The aim of this research was to assess the changes in SDR and other parameters of bone density distributions within the skull and to demonstrate that not only SDR can be increased, but overall enhanced bone mass can be built through alendronate treatment. This, in turn, enables a successful repeat thalamotomy.

\section{Materials and Methods}

\subsection{Study design}
\label{subsec_study}

The study is retrospective and based on a series of clinical cases. 

Five clinical cases of patients who underwent an unsuccessful initial attempt at thermal ablation using focused ultrasound on the ExAblate Neuro 4000 system (InSightec, Israel) have been examined. Clinical treatments were conducted at the V.S. Buzaev International Medical Center in Ufa, Russia, from 2021 to 2025.

In four out of five patients, the SDR initially exceeded the 0.35 threshold, indicating a reasonable candidate for the treatment. However, in three cases, the target temperature of 55°C,  necessary for effective tissue ablation was not achieved. Additionally, in two cases, the procedure had to be halted due to severe pain experienced during sonication. This highlights the challenges encountered during the procedure despite the satisfactory SDR.

All patients were referred to an endocrinologist for comprehensive therapy aimed at strengthening skull bone tissue. During a period of 6 to 12 months, they received a treatment regimen that included alendronic acid, vitamin D, and calcium supplements.

\subsection{Patients}
\label{subsec_patients}

This study analysed cases of five patients (labelled A–E) aged from 45 to 65 years, with a mean age of 55.4 ± 8.2 years at the start of treatment. They underwent MRgFUS non-invasive ablation surgery with diagnoses of Parkinson’s disease (PD), essential tremor (ET), and dystonia. A detailed list of patients, including gender, age, diagnosis, and treatment progression, is provided in Table~\ref{Table1}.

In all cases, the initial attempt at ultrasound ablation was unsuccessful. For patients A and D, the procedure was interrupted due to severe pain when the temperature reached the target of 55°C. For the remaining patients (B, C, and E), the target temperature and effective ablation of the target area could not be achieved despite multiple sonication attempts. Clinical guidelines recommend treating patients with an SDR > 0.35 (or 0.40 in some countries \cite{wang2018transcranial, ng2024magnetic}), which was met for four out of five patients. However, patient B had an SDR of 0.32.

Following the unsuccessful ultrasound ablation attempts, patients were consulted by an endocrinologist and began taking medications for osteoporosis (alendronic acid, cholecalciferol, and \textit{Calcium D3 Nycomed} by \textit{Takeda}, Russia) aimed at strengthening bone tissue. The treatment duration ranged from six months to one year, with specific drug regimens and dosages detailed in Table~\ref{Table1}. After completing the course of treatment, a second MRgFUS ablation surgery was performed at least one year after the initial procedure, and it was successful in all cases.

\begin{table*}[]
\small
\begin{center}
\begin{tabular}{cp{2.5cm}cp{5cm}p{2.5cm}p{3.5cm}}%{lllllll}
Patient & Age (1st-2nd treatment) / Gender & Disease  & Treatment & Duration & Special notes                               \\
\hline
A       & 65-67 M      & PD       & Alendronic acid 70 mg once a week, simultaneously "Calcium D3 Nycomed" 500/200   mg twice a day,
and cholecalciferol 7000 IU/day for 2 months, then 2000   IU/day. & 180 days  & Pain during the first surgery 7 points on VAS \\
\hline
B       & 58-61 F      & ET       & Alendronic acid 35 mg once a week, simultaneously "Calcium D3 Nycomed" 500/200   mg twice a day, and cholecalciferol 3000 IU/day.                                  & 180 days  &   Ablation temperature was not achieved during the first surgery                                          \\
\hline
C       & 49-51 M      & ET       & Alendronic acid 35 mg once a week, simultaneously "Calcium D3 Nycomed" 500/200   mg twice a day, and cholecalciferol 3000 IU/day.                                  & 180 days  &  Ablation temperature was not achieved during the first surgery                                           \\
\hline
D       & 45-47 M      & Dystonia & Alendronic acid 70 mg once a week, simultaneously "Calcium D3 Nycomed" 500/200   mg twice a day, and cholecalciferol 2000 IU/day.                                  & 365 days (control CT after 180 days with no change in SDR) & Pain during the first surgery 9 points on VAS \\
\hline
E       & 60-61 F      & PD       & Alendronic acid 70 mg once a week, simultaneously "Calcium D3 Nycomed" 400/200   mg twice a day, and cholecalciferol 2000 IU/day for 2 months.                     & 180 days  & Ablation temperature was not achieved during the first surgery                                         
\end{tabular}
\end{center}
\caption{Patient demographics, treatment regimen, and clinical notes}\label{Table1}
\end{table*}

\subsection{Statistical Analysis}
\label{subsec_stats}

Statistical analysis has been performed to determine whether the maximum temperature at the focal spot increased during treatment when comparing between the initial and repeat MRgFUS surgery. All sonications for five patients in \textit{Treat} mode were analysed, and two groups of maximum temperatures during sonications — corresponding to the unsuccessful first and the successful second surgery— were compared using the Wilcoxon rank-sum left-sided test. The maximum temperatures recorded throughout the procedure were also analysed for each patient during both the unsuccessful first and the successful second surgery. This paired data was further evaluated using the Wilcoxon signed-rank left-sided test.

\subsection{CT Imaging Protocol}
\label{subsec_ct}

CT scans were acquired using protocols optimised for high-contrast bone imaging. It is important to note that pre- and post-alendronate treatment CT scans were not always performed using the same scanning mode. The imaging parameters, including the scanner model, convolution kernel, filter type, kVp, and tube current, were documented for both pre- and post-treatment scans. These details are summarised in Table~\ref{Table2} to ensure transparency and reproducibility of the imaging protocol.

\begin{table*}[ht]
\normalsize
\begin{center}
\begin{tabular}{clllll}
\multicolumn{1}{l}{Patient} & CT scanner                                                & Filter type                         & Convolution   kernel             & Tube   current         & KVP                         \\
                              & \cellcolor[HTML]{D9D9D9}Siemens SOMATOM   Definition AS+  & \cellcolor[HTML]{D9D9D9}flat        & \cellcolor[HTML]{D9D9D9}H60s     & \cellcolor[HTML]{D9D9D9}176 & \cellcolor[HTML]{D9D9D9}120 \\
\multirow{-2}{*}{A}           & Siemens SOMATOM Definition AS+                            & flat                                & H60s                             & 176                         & 120                         \\
                              & \cellcolor[HTML]{D9D9D9}Siemens SOMATOM   Emotion 16      & \cellcolor[HTML]{D9D9D9}1           & \cellcolor[HTML]{D9D9D9}H60s     & \cellcolor[HTML]{D9D9D9}54  & \cellcolor[HTML]{D9D9D9}110 \\
\multirow{-2}{*}{B}           & Siemens SOMATOM Emotion 16                                & 0                                   & H60s                             & 154                         & 120                         \\
                              & \cellcolor[HTML]{D9D9D9}Siemens SOMATOM   Definition Edge & \cellcolor[HTML]{D9D9D9}flat        & \cellcolor[HTML]{D9D9D9}H60s     & \cellcolor[HTML]{D9D9D9}201 & \cellcolor[HTML]{D9D9D9}100 \\
\multirow{-2}{*}{C}           & Siemens SOMATOM Definition Edge                           & flat                                & H60s                             & 141                         & 120                         \\
                              & \cellcolor[HTML]{D9D9D9}Siemens SOMATOM   Sensation 16    & \cellcolor[HTML]{D9D9D9}0           & \cellcolor[HTML]{D9D9D9}H60s     & \cellcolor[HTML]{D9D9D9}154 & \cellcolor[HTML]{D9D9D9}120 \\
\multirow{-2}{*}{D}           & Siemens SOMATOM Sensation 16                              & 0                                   & H60s                             & 154                         & 120                         \\
                              & \cellcolor[HTML]{D9D9D9}GE Medical Systems   BrightSpeed  & \cellcolor[HTML]{D9D9D9}head filter & \cellcolor[HTML]{D9D9D9}BONEPLUS & \cellcolor[HTML]{D9D9D9}151 & \cellcolor[HTML]{D9D9D9}120 \\
\multirow{-2}{*}{E}           & GE Medical Systems BrightSpeed                            & head filter                         & BONEPLUS                         & 151                         & 120                        
\end{tabular}
\end{center}
\caption{CT imaging protocols for 5 patients before (grey) and after (white) the alendronate treatment session.}\label{Table2}
\end{table*}

\subsection{Image Registration and Analysis}
\label{subsec_reg}

\begin{figure}[htbp]
\centering
\includegraphics[width=0.5\textwidth]{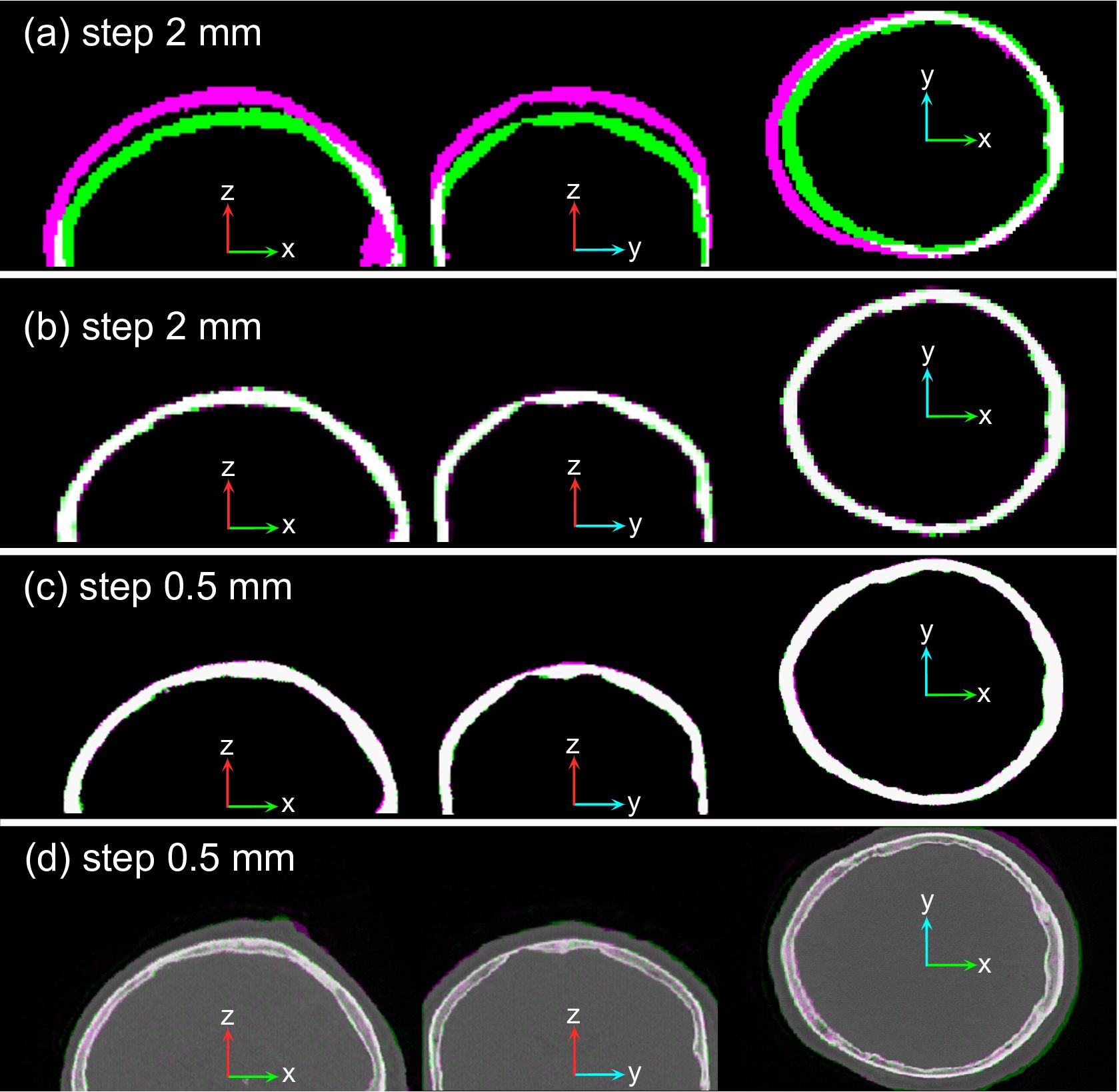}
\caption{Process of CT image registration before and after alendronate treatment. (a) Initial skull masks with low spatial resolution; (b) registration of masks with low resolution; (c) registration of masks with high resolution; (d) registration of CT images with high resolution.}
\label{fig:registration}
\end{figure}

To analyse changes in bone density distribution, CT image registration was performed. Original CT scans, initially in DICOM format, were converted to NRRD format. The lower part of the head, which was not exposed to ultrasound during MRgFUS procedure, was removed from the files. The images were centered, empty margins were cropped to minimise the array size, and the scans were interpolated onto an equidistant grid with a spatial step of 0.5 mm using the nearest-neighbour method \cite{robertson2018effects}. This method was chosen to preserve HU values in voxels during rotation and translation of the CT scan during registration, ensuring a more accurate assessment of overall skull density changes. A two-step automatic image registration process was then performed in the Matlab environment, described in more detail in \cite{krokhmal2025comparative}. The graphical representation of the registration steps is shown in Fig.~\ref{fig:registration}.

After precise alignment of the two CT scans, a skull bone mask was generated in Matlab using a threshold value of 200 HU. A skull mask represented a binary 3D array with 1 for skull tissue and 0 for others. Numerical analysis was then performed only within the identified mask area.

Alendronate treatment leads to trabecular bone densification, which alters the brightness of specific skull regions in CT images. To quantitatively analyse these changes, density histograms (in HU units) were constructed for all skulls before and after the treatment course, showing the percentage of voxels with a given X-ray density value. In two cases, pre- and post-treatment CT scans were acquired with different scanner settings (Table 2). Furthermore, as noted in \cite{goldman2007principles}, increasing the radiation dose reduces image noise but can reduce subject contrast as well. Therefore, direct comparison of the HU changes in specific bone regions alone would not provide a reliable quantitative assessment of density changes due to scanner calibration and settings differences \cite{webb2018measurements}. 

Additional calibration of the images has been performed to quantitatively evaluate the changes and re-calibrated histograms were generated to reflect the volumetric fraction of voxels as a function of HU. Due to variations in CT scanner calibrations \cite{free2018effect}, each post-treatment CT scan was normalised to the pre-treatment scan by multiplying the HU values by a scaling factor so that the histogram regions above 1500 HU were aligned, and the HU value for water (0 HU) did not change. This approach was based on the following considerations: it is known \cite{roschger1997mineralization, roschger2001alendronate} that osteoporosis medications primarily affect trabecular bone due to its higher metabolic rate. In trabecular bone, tissue resorption slows down, and the trabecular structure is preserved and strengthened, while the structure and density of cortical bone remain largely unchanged. Therefore, we assumed that the primary histogram changes would occur in the range between 300 HU, that corresponds to the average trabecular bone density, and 1200 HU, that corresponds to the average cortical bone density \cite{treece2018cortical}, while the densest skull regions above 1500 HU would not undergo further significant densification.

To assess local changes in the physical density of skull regions, HU values were converted to physical density units according to
\cite{schneider1996calibration}.

\section{Clinical Cases}

\subsection{Patient A, Male, 68 years old}  

Patient with a  PD, tremor dominant form, predominantly affecting the right extremities, Stage 3 \cite{hoehn1967parkinsonism}.  Neurological status: moderate resting tremor in both upper limbs, exacerbated by stress; tremor of the lower jaw; postural, kinetic, and intentional tremor in the upper limbs, more on the right side; marked extrapyramidal symptoms on the right side; propulsion, akinesia on both sides, and bradykinesia.  

The disease was diagnosed in 2012, and the patient was on medication therapy "PC Merz" (200 mg, twice daily) and  "Nakom" (Levodopa 250 + Carbidopa 25 mg, three times daily). Given the persistence of symptoms despite treatment, the patient was referred for MR-guided FUS treatment.  

Preoperative CT showed an SDR of 0.42.  
As there were no contraindications, the procedure was initiated but prematurely terminated due to significant pain (VAS 7/10) in the treatment area.  

A multidisciplinary team (neurosurgeon, neurologist, endocrinologist, and cardiologist) prescribed bisphosphonate therapy to improve bone density before the second intervention (Table~\ref{Table1}).

After six months, repeated CT showed an SDR increase to 0.48, and a second MRgFUS procedure was successfully performed.  
Follow-up showed a significant improvement in the resolution of tremor, bradykinesia, and hypokinesia. No adverse effects from bisphosphonate therapy were observed.    

\subsection{Patient B, Female, 61 years old}  

Patient with a primary diagnosis of ET, familial case, involving the upper limbs and head. She suffered significant impairment of social and daily adaptation. The tremor began at the age of 40. Initially, the tremor was situational—worsening under stress. By the age of 50, the tremor had significantly intensified. Medical history includes type 2 diabetes mellitus, postoperative hypothyroidism, and thyroidectomy. 

In 2020, the patient underwent non-invasive thalamotomy of the VIM nucleus. The procedure lasted 2 hours and 55 minutes, with intraoperative headache (3-5 on the VAS scale), managed with non-steroidal anti-inflammatory drugs (NSAIDs). Following the focused ultrasound treatment in 2020, the tremor in the right hand did not bother her for 4 months, after which it gradually began to return. 

Given the pre-operative SDR of 0.32 and the unstable clinical effect, the multidisciplinary team decided to propose a repeat surgical intervention, with preliminary adjustment of therapy aimed at improving the ultrasound conductivity of the bone tissue (Table~\ref{Table1}). After 6 months of therapy, a repeat computed tomography scan of the skull was performed, showing an SDR of 0.41. 

The patient underwent a repeat MRgFUS intervention: unilateral non-invasive thalamotomy of the VIM. At the 3-month follow-up, an impressive reduction in kinetic tremor in the arms—up to 80\% was observed.  However, a significant intention tremor remains, which partially limits daily activities

\subsection{Patient C, Male, 52 years old}  

Patient with ET. The hand tremor first appeared at the age of 20. In 2013, he was diagnosed with essential tremor and propranolol was prescribed with a positive effect. Clonazepam was also prescribed but proved ineffective. Over the past 5 years, the tremor has progressively worsened.

He underwent a unilateral non-invasive MRgFUS thalamotomy of the left VIM nucleus. Prior to the intervention, the SDR was 0.39. The procedure lasted 2 hours and 40 minutes, with a positive outcome: 90\% reduction in kinetic postural tremor of the right hand and 70\% reduction in intention tremor. The patient was satisfied with the clinical outcome, and the decision was made to halt the procedure. The perioperative period was uneventful; however, the tremor recurred after 3 months, and the patient expressed a desire to repeat the operation for further reduction of the intention tremor.  

The multidisciplinary team has made a pragmatic decision to start therapy as outlined in Table~\ref{Table1}, before considering another surgical intervention. This therapy increased the SDR to 0.46, and the patient underwent a repeat operation. The procedure lasted 1 hour and 10 minutes, with a positive outcome: 95\% reduction in kinetic postural tremor of the right hand and 90\% reduction in intention tremor.

\subsection{Patient D, Male, 47 years old}  

Patient with a primary diagnosis of focal craniocervical persistent idiopathic sporadic dystonia with onset in adulthood and a progressive course. Neurological status: Irregular, intermittent, medium-amplitude hyperkinesis with retrocollis and rightward laterocollis. The patient can turn the head 90 degrees to the right and 70 degrees to the left without corrective movements.  During walking, laterocollis and hyperkinesis persist. Occasional involuntary movements in the left upper limb occur during emotional stress.  

The patient underwent surgery on the right sternocleidomastoid muscle several years ago in another hospital, with a positive effect lasting 12 months, after which the aforementioned symptoms returned. The patient received numerous times botulinum toxin therapy, with transient positive effects.

Given the absence of contraindications (SDR = 0.36), an attempt was made to perform unilateral non-invasive treatment using focused ultrasound on the left side. During the procedure, the patient reported pain in the intervention area (8-9 on the VAS scale), leading to the decision to halt the operation. 

To improve skull bone density, bisphosphonate therapy was prescribed as mentioned in (Table~\ref{Table1}). After 12 months of therapy, a repeat skull CT scan was performed, showing an SDR of 0.41. The patient then underwent a repeat MRgFUS treatment. At the time of writing this article, neurological symptoms had not recurred.

\subsection{Patient E, Female, 61 years old}  

Patient was diagnosed with PD, treated with Levodopa and Pramipexole. She underwent thalamotomy under the guidance of focused ultrasound. The SDR before the first procedure was 0.39. During the operation, the target temperature parameters were not fully achieved. Postoperatively, there was a positive effect with the disappearance of tremor, but bradykinesia persisted, prompting the planning of a repeat surgical intervention after correcting calcium-phosphate metabolism according to Table~\ref{Table1}.

After the therapy, the SDR decreased to 0.37. Despite this, a repeat operation was performed, during which the target parameters were successfully achieved. The patient reported clinical improvement, including regression of bradykinesia. However, the therapy was continued due to the need for surgery on the opposite side. Additionally, densitometry revealed osteopenia in the patient.

\section{Results}

\subsection{Temperature Rises}
\label{subsec_temperature}

\begin{table*}[htbp!]
\normalsize
\begin{center}
    \begin{tabular}{ccccccc}
    \multicolumn{1}{l}{Patient} & $N_{s}$             & SDR                          & Average Energy,   kJ             & Average T, C                     & Max Energy, kJ                    & Max T, C                   \\
    \hline
                                  & \cellcolor[HTML]{D9D9D9}9 & \cellcolor[HTML]{D9D9D9}0.42 & \cellcolor[HTML]{D9D9D9}30.1±9.3 & \cellcolor[HTML]{D9D9D9}51.0±3.6 & \cellcolor[HTML]{D9D9D9}36.0      & \cellcolor[HTML]{D9D9D9}55 \\
    \multirow{-2}{*}{A}           & 8                         & 0.48                         & 26.1±9.8                         & 56.0±3.0                         & 30.1-36.1                         & 60                         \\
    \hline
                                  & \cellcolor[HTML]{D9D9D9}8 & \cellcolor[HTML]{D9D9D9}0.32 & \cellcolor[HTML]{D9D9D9}33.9±4.1 & \cellcolor[HTML]{D9D9D9}53.6±2.6 & \cellcolor[HTML]{D9D9D9}30.1      & \cellcolor[HTML]{D9D9D9}58 \\
    
    \multirow{-2}{*}{B}           & 11                        & 0.41                         & 35.9±0.5                         & 53.7±4.2                         & 34.2                              & 60                         \\
    \hline
                                  & \cellcolor[HTML]{D9D9D9}6 & \cellcolor[HTML]{D9D9D9}0.39 & \cellcolor[HTML]{D9D9D9}20.3±7.1 & \cellcolor[HTML]{D9D9D9}59.8±1.3 & \cellcolor[HTML]{D9D9D9}22.3 & \cellcolor[HTML]{D9D9D9}62 \\
    \multirow{-2}{*}{C}           & 3                         & 0.46                         & 10.9±2.5                         & 62.0±1.7                           & 8.0-12.4                          & 63                         \\
    \hline
                                  & \cellcolor[HTML]{D9D9D9}6 & \cellcolor[HTML]{D9D9D9}0.37 & \cellcolor[HTML]{D9D9D9}24.1±7.0 & \cellcolor[HTML]{D9D9D9}54.3±2.7 & \cellcolor[HTML]{D9D9D9}30.6      & \cellcolor[HTML]{D9D9D9}58 \\
    \multirow{-2}{*}{D}           & 10                        & 0.40                         & 20.7±8.6                         & 55.1±4.0                         & 34.8                              & 60                         \\
    \hline
                                  & \cellcolor[HTML]{D9D9D9}9 & \cellcolor[HTML]{D9D9D9}0.39 & \cellcolor[HTML]{D9D9D9}31.0±8.7 & \cellcolor[HTML]{D9D9D9}51.7±2.9 & \cellcolor[HTML]{D9D9D9}30.0-36.0 & \cellcolor[HTML]{D9D9D9}54 \\
    \multirow{-2}{*}{E}           & 3                         & 0.37                         & 30.9±2.7                          & 57.3±1.2                        & 29.3                              & 58                        
    \end{tabular}
\end{center}
\caption{Changing parameters during MRgFUS sonications for pre(grey) and post(white) alendronate treatment.}\label{Table3}
\end{table*}

The main result of this study is the clinical success of the repeat procedure, which means that the target temperature of 55 ° C was achieved for all patients. Each patient underwent multiple sonications – MRregFUS treatment trials, when different power levels, exposure durations, and target points could be selected (see Table~\ref{Table3}).  $N_s$ represents the number of sonications per individual patient.

Table~\ref{Table3} presents the changes in the temperature metrics for the patients during the first and second surgeries. Specifically, it shows the average maximum temperatures per sonication for each patient in both procedures (shown as Average T) (number of sample elements for statistical analysis was N = 38 and N = 34, respectively). In each case, the focal temperature increased during the repeated procedure. As seen in Table~\ref{Table3}, during the repetition procedure, the average maximum heating temperature between sonications increased for each patient. The average maximum focal temperature for all sonications before alendronate treatment was 53.6 ± 4.0° C, whereas after treatment, it increased to 55.7 ± 4.1° C. This temperature increase was statistically significant (p = 0.018).  The smallest change was observed in patient B (from 53.6 ± 2.6° C to 53.7 ± 4.2° C), while the highest temperature increase was recorded in patient E (from 51.7 ± 2.9° C to 57.3 ± 1.2° C).  

We also analysed the change in the maximum focal temperatures per patient throughout the procedure, shown as Max T in the last column of Table~\ref{Table3}, (N = 5). During the first surgery, the mean maximum temperature was 57.0 ± 2.4° C, while in the second procedure, it increased significantly to 60.2 ± 1.8° C (p = 0.031). This increase was sufficient to ensure the success of the repeat procedure.

The average energy required to achieve the desired temperature during sonication (shown as Average Energy in Table~\ref{Table3}, N = 38 and 34, before and after alendronate treatment, respectively) decreased slightly, from 28.7 ± 8.6 kJ to 26.7 ± 10.0 kJ. For Patient C, it was possible to reduce the energy by half after alendronate treatment while still reaching the target temperature of 63° C in the focal area. In contrast, for Patient D, the energy was increased during the repeat procedure, but unlike the first attempt prior to alendronate treatment, this did not result in any pain.

\begin{figure*}[htbp]
    \centering   
    \includegraphics[width=1\textwidth]{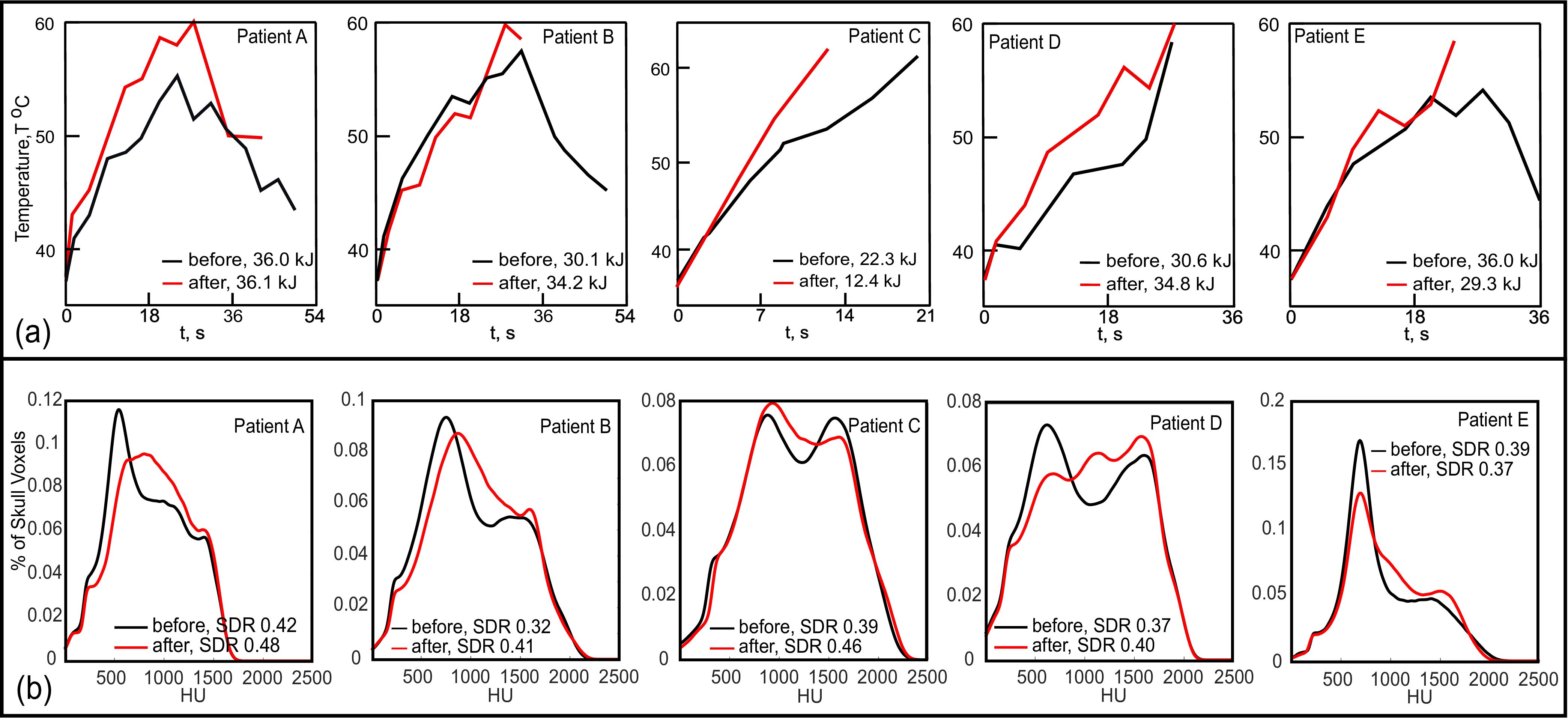}
    \caption{(a) The temperature during the MRgFUS sonication with the maximum heating for each patient, (b) normalized histograms of CT scans of the sonicated skull region. Black and red curves represent cases before and after the alendronate therapy.}
    \label{fig:t_rise}
\end{figure*}

Fig.~\ref{fig:t_rise} (a) shows the temperature rise curves for each patient, corresponding to the maximum temperature achieved during the entire procedure. The black curves represent the temperature changes before alendronate treatment, while the red curves depict the changes after treatment. The energy levels used in each case are also indicated. As seen in the graphs, a higher focal temperature was achieved in all cases. For Patients C and E, this was accomplished in a shorter time and with lower energy. For Patients A and D, the temperature increased during the repeat procedure, with the same or higher energy levels. Unlike the initial treatment, this did not cause any pain.

\subsection{SDR Changes}
\label{subsec_sdr}

\begin{figure}[t]
    \centering   
    \includegraphics[width=0.5\textwidth]{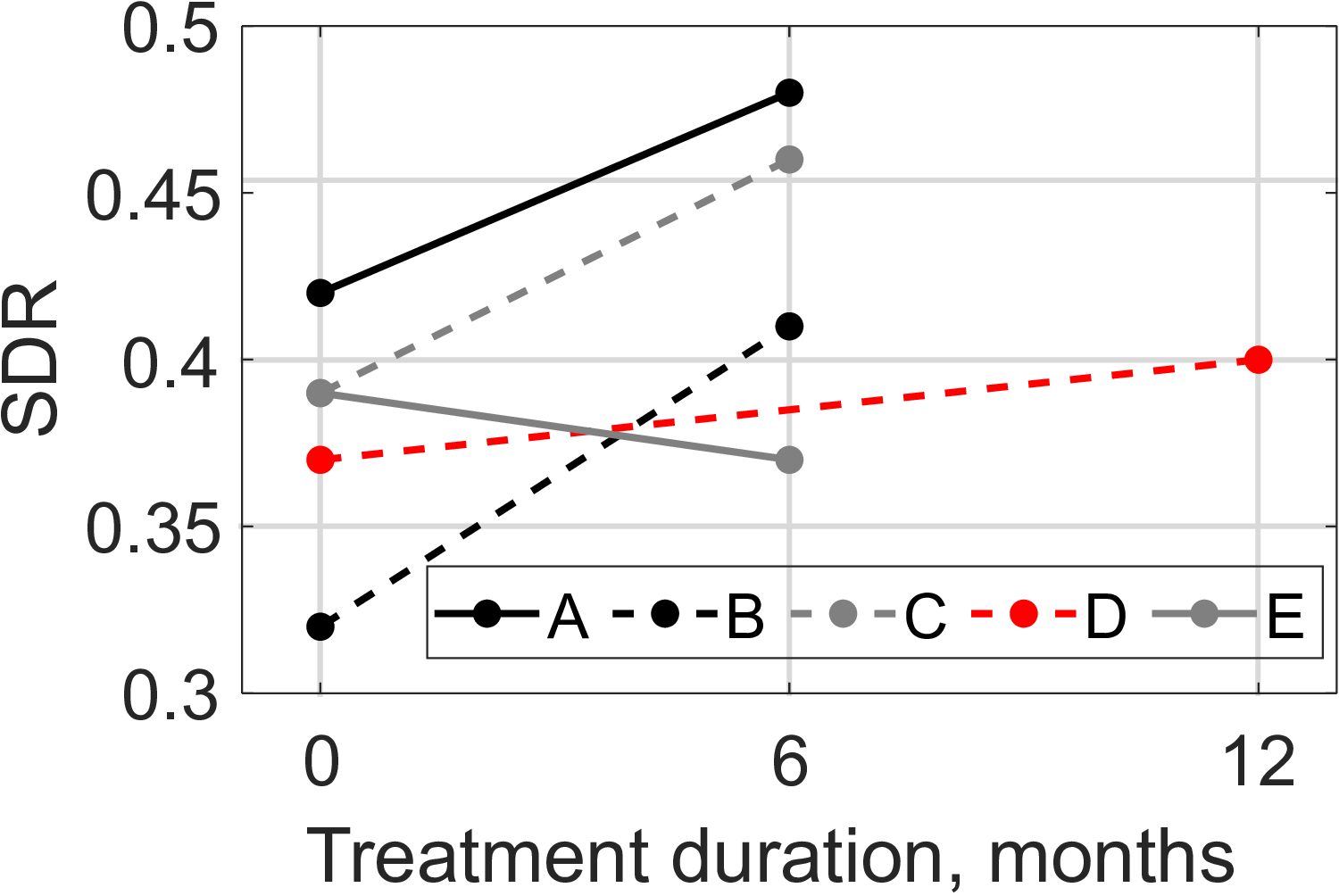}
    \caption{Change in SDR  during the alendronate therapy.}
    \label{fig:sdr}
\end{figure}

Initially, 4 out of 5 patients had an SDR above the threshold value of 0.35 (ranging from 0.32 to 0.42, with an average of 0.378 ± 0.037). After 6–12 months of therapy, the SDR increased in 4 out of 5 patients (reaching an average of 0.424 ± 0.045, ranging from 0.37 to 0.48). For Patient E, the SDR decreased from 0.39 to 0.37 (Table~\ref{Table3}), but this did not hinder achieving the target temperature rise during the repeat procedure. The dynamics of SDR changes is illustrated in Fig.~\ref{fig:sdr}.

\begin{figure*}[htbp]
    \centering   
    \includegraphics[width=0.9\textwidth]{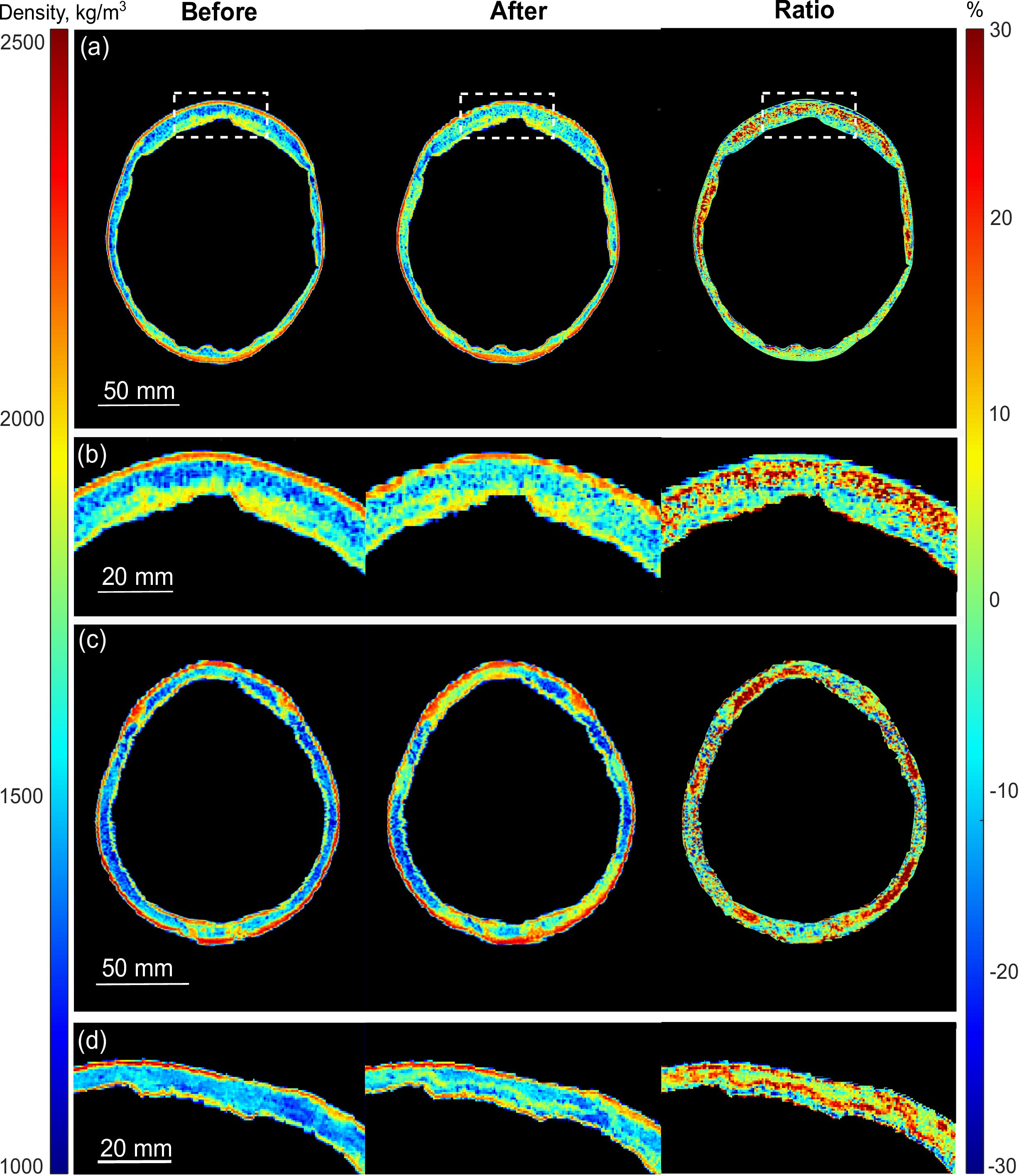}
    \caption{Density distributions within the compared skull sections before and after alendronate treatment (columns 1 and 2), as well as the percentage change in density (column 3): (a) and (b) are a larger and more detailed image for Patient B, the area of figure (b) is indicated in figure (a) by a white dashed line; (c) and (d) correspond to Patients D and E.}
    \label{fig:ratios}
\end{figure*}

\begin{figure*}[htbp]
    \centering   
    \includegraphics[width=1\textwidth]{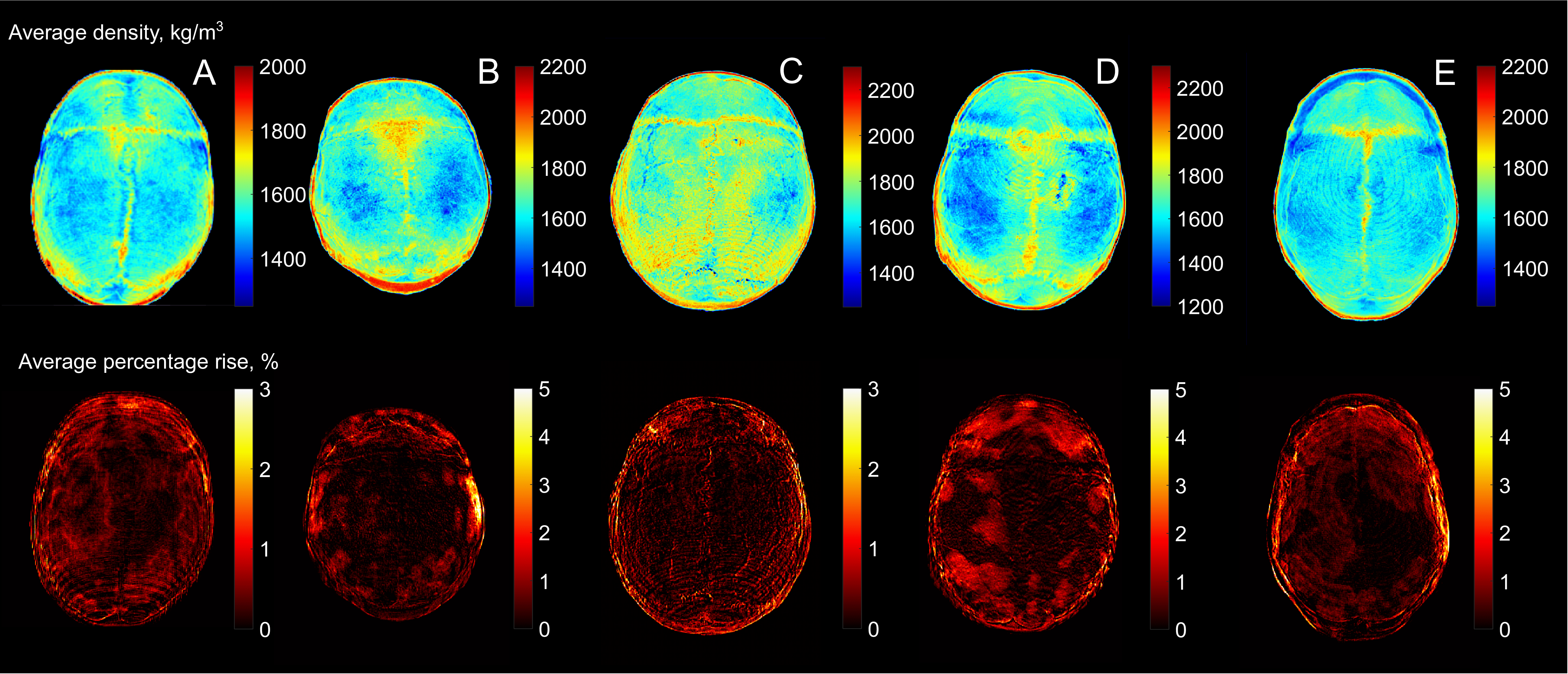}
    \caption{Skull density before treatment and areas of densification on skulls; skulls are labelled (A – E) according to the numbering of patients.}
    \label{fig:surface}
\end{figure*}

\subsection{Histograms Changes}
\label{subsec_histograms}

Fig.~\ref{fig:t_rise} (b) shows the calibrated histograms of CT scans of the sonicated skull region for each patient before and after treatment. As seen in the graphs, in every case, there is a decrease in the number of voxels corresponding to lower HU values, while the proportion of voxels with higher X-ray density increases.

To varying degrees, a shift toward higher HU values is observed in the first peak, which falls within the 500–1000 HU range. This peak likely corresponds to the widespread densification of trabecular bone tissue, a change that is particularly noticeable in patients A and B, as shown in Fig.~\ref{fig:t_rise} (b).

The second peak, located around 1500 HU and presumably representing the cortical bone mass, remains unchanged across all cases. However, in four out of five cases, the peak amplitude is higher, which can be interpreted as an increase in the number of voxels with this density, indicating thickening of the cortical layer in correspondence with \cite{seeman2010microarchitectural}.

Additionally, in all cases, the transition between the first and second peaks becomes smoother, suggesting an increase in the number of voxels with intermediate values around 1000 HU. This indicates a more gradual density transition within the skull bone, which likely leads to reduced acoustic wave reflection.

Overall, the mean HU value of the entire bone volume increased in all cases, ranging from 2.5\% (patient C) to 6.2\% (patient A).

\subsection{Decrease of porosity}

To evaluate which areas of the skull are most affected by bisphosphonate treatment, CT images before and after treatment were registered, as described in Sec.~\ref{subsec_reg}. The CT scans, measured in HU, were calibrated and then converted into a skull bone density distribution map according to \cite{schneider1996calibration}. Then these maps were compared layer by layer for each patient. The percentage change in skull density within each voxel was also calculated. Fig.~\ref{fig:ratios} shows the most illustrative examples of changes in skull density for different patients. The images reveal that the most significant changes occur in the trabecular bone. For instance, in Fig.~\ref{fig:ratios} (a, b), the most porous areas of the bone, shown in blue, exhibit the most noticeable densification. In the third column, these areas are marked in red, indicating the regions with the highest percentage increase in local density. 

It is also evident that the densification of the trabecular bone is not uniform across the entire skull. For example, in Fig.~\ref{fig:ratios} (c), the skull bone is quite porous throughout the region before the treatment, but significant densification occurred in specific areas of the frontal, temporal, and parietal bones, highlighted in red in the third column. Another interesting case is that of patient E, whose temporal bone slice is shown in Fig.~\ref{fig:ratios} (d). Here, the increase in trabecular bone density did not occur uniformly throughout the porous volume but rather in a central layer within the bone.

An analysis was also conducted to determine which regions of the skull exhibit the most significant density changes in response to alendronate treatment. Fig.~\ref{fig:surface} presents surface maps showing the density before alendronate treatment and the percentage change in bone density after the treatment, both values were averaged by the vertical axes of the skull for better visual representation.  As seen in the images, the densification is most pronounced in the frontal bone and the lower part of the parietal bone, while changes in the crown are the least noticeable. In patients A, B, and C, the densification appears more localized, whereas in patients C and D, it is more evenly distributed throughout the region. It is also notable that the skull regions with the highest density are mostly unchanged, while some regions of low density tend to increase. 

\section{Discussion}

This study demonstrate that the use of alendronate appears to enhance bone density distributions within the skull, particularly in the trabecular bone, leading to a more uniform density distribution and, consequently, improved ultrasound transmission. As discussed in Sec.~\ref{subsec_histograms}, histogram analysis before and after the alendronate therapy reveals a shift of voxels, primarily those within the 200–1000 HU range, towards higher HU values. This indicates active mineralisation of the trabecular bone, consistent with clinical observations. According to   \cite{zebaze2005cortical}, treatment effects treatment effects are typically more prominent in trabecular bone due to its higher surface-to-volume ratio and greater capacity for remodelling compared to the cortical bone. Such areas of trabecular bone are marked in Fig. 4 as regions showing the most pronounced local densification. 

Changes in trabecular bone structure following prolonged alendronate treatment have been previously reported \cite{roschger2001alendronate}. In that study it was found that mineral distribution within individual trabeculae became more uniform after 2–3 years of continuous treatment, a result that aligns with findings from a preclinical study in minipigs, where cancellous bone matrix showed similar homogenisation of mineralisation alongside reduced bone turnover 
\cite{roschger1997mineralization}. Moreover, a modest increase in average mineral density in trabecular bone was observed after three years of treatment.

Simultaneously, changes in cortical bone were also evident. In 4 out of 5 patients considered in the current study, an increase in histogram amplitude was observed within the 1000–2000 HU range, suggesting a rise in the number of voxels with HU values characteristic of cortical bone. This implies cortical thickening, which is particularly apparent in Patient E (see Fig.~\ref{fig:t_rise} and Fig.~\ref{fig:ratios} (d)). Visual inspection shows not only the densification of the trabecular layer but also cortical thickening and densification. Notably, Patient E exhibited a lower SDR after the alendronate treatment, despite positive structural changes. This agrees with the findings of \cite{seeman2010microarchitectural}, who reported that reduced bone remodelling due to antiresorptive therapy leads to increased cortical thickness. A plausible explanation is that reduced cortical porosity—through infilling with mineralised bone—increases the overall mineralised area.

Thus, alendronate treatment results in a more uniform distribution of bone density, consistent with findings by  \cite{roschger2001alendronate}. This is reflected in the density histograms of the skull (Fig.~\ref{fig:t_rise} (b)), where the difference between low and high HU voxel counts is reduced, and peaks appear smoother. These changes suggest, in addition to the SDR value, 3D density distribution within the skull, represented as histograms, may provide additional insight into the acoustic permeability of the skull. For instance, \cite{marsac2017ex} presented histograms of sound speed (linearly related to HU) for ex vivo skull samples and reported ultrasound transmission coefficients. Skulls with histograms peaking at higher sound speed values—and with fewer low-speed voxels—had better transmission. This observation is consistent with our findings, suggesting that a metric based on the density distribution within the skull bone irradiated by MRgFUS field could be a valuable tool to compare transmission efficiency before and after alendronate treatment.

The need for supplementary metrics beyond SDR planning surgical HIFU treatments is evident. Although in our study the SDR increased in 4 of 5 patients, one patient with a decreased SDR still underwent successful treatment. This implies that in clinical practice, in borderline cases, additional metrics may support better-informed decisions. SDR is widely used as a threshold metric in clinical settings, with a common cut-off of 0.4 \cite{wang2018transcranial}. However, several studies have questioned its reliability in accurately representing ultrasound transmission efficiency \cite{tsai2021distribution}. While SDR values above 0.45 are generally associated with better transmission, successful treatment remains possible at lower values \cite{d2019impact}. Patients with SDR < 0.4 may require increased energy delivery, but treatment outcomes are not necessarily inferior \cite{boutet2019relevance}. Recent data show that patients with SDR < 0.40 can be safely and effectively treated with MRgFUS, although there is a potentially higher risk of treatment failure and intraoperative discomfort \cite{ng2024magnetic}.

It is important to note that SDR values may vary between different CT scanners and manufacturers. From our experience, the SDR threshold appears to depend on scanner-specific parameters. To accurately track changes in a given patient over time, the same CT scanner and acquisition protocol should be used. Alternatively, normalisation is essential, which can be performed, for example, by using an electron density calibration phantom \cite{constantinou1992electron} for the CT scanner or by the histogram normalisation method, proposed in this work in Sec.~\ref{subsec_reg}. However, substantial changes in cortical bone — as seen in Patient E — may complicate histogram normalisation, particularly at the higher HU values. Fortunately, in our study, scans for this patient were performed using identical imaging parameters, allowing us to confidently attribute observed changes in histogram shape to physical effects rather than to imaging artefacts.

Given the limitations of SDR, alternative metrics have already been proposed to better correlate with clinical outcomes. For example,  \cite{tsai2021distribution} introduced a customised SDR metric based on the ratio of average trabecularto cortical bone intensity, using a developed skull segmentation algorithm. However, to date, the only relatively reliable method to predict MRgFUS treatment efficacy remains numerical modelling of ultrasound beam propagation through the skull, incorporating aberration correction techniques, as demonstrated by \cite{mcdannold2019elementwise}.

One limitation of the proposed approach to enhancing ultrasound permeability is the variability in patient response to alendronate. Most patients in our cohort were over 60 years old, a demographic often affected by osteopenia. In such cases, alendronate tends to mineralise regions with the greatest deficit—typically trabecular bone. For example, in our experience, one osteopenic patient with a low baseline SDR showed no post-treatment changes in SDR, histogram profile, or visual CT appearance, despite receiving a year-long course of alendronate. However, densitometry of the lumbar spine and femur showed significant improvements, indicating a systemic response. A longer treatment duration, or switching to a different medication, may prove more effective in mineralising the skull. For instance, studies have shown that denosumab suppresses bone remodelling more strongly than alendronate, leading to greater increases in areal bone mineral density and cortical layer thickness  \cite{seeman2010microarchitectural}. These differences may reflect the distinct effects of each drug on bone microarchitecture and strength.

\section{Conclusions}

This retrospective study demonstrates that alendronate treatment for 6 - 12 months yields promising results in improving the efficacy of MRgFUS ablation of brain tissue by improving the acoustic permeability of the skull, without any reported side effects from the medication. These findings suggest that bisphosphonates could serve as a beneficial preparatory treatment prior to MRgFUS interventions for patients with low skull density, by promoting mineralization and contributing to a more uniform density distribution. These changes are confirmed by CT histogram analysis and visual CT inspection. Importantly, this effect is not always captured by the commonly used SDR metric, which may not reliably reflect true transmission efficiency, particularly in the borderline or atypical cases. Our findings suggest that density distribution metrics can provide valuable supplementary information during pre-treatment planning for MRgFUS procedures. However, while the results are encouraging, further research is essential to validate these outcomes and better understand the long-term implications of bisphosphonate therapy in this context. Continued investigation will be vital in establishing a robust clinical framework for integrating this treatment strategy into standard care protocols.

%\appendix
%\section{Example Appendix Section}
%\label{app1}
\medskip

%\bibliography{thebibliography}
%\bibliography{sampbib}

\bibliographystyle{plain}

\end{document}